\documentclass[a4paper,12pt]{article}
\usepackage[margin=1.2in]{geometry}
\usepackage{graphicx}
\usepackage{subfigure}
\usepackage[square,numbers]{natbib}
\usepackage{lipsum}
\usepackage{caption}
\usepackage{pdfpages}
\usepackage{amsmath}
\usepackage{multicol}
\usepackage{multirow}
\usepackage{float}
\usepackage{url}
\usepackage{appendix}
\usepackage{bm}
\usepackage{amssymb}

\usepackage{mwe}

\usepackage{fullpage, comment}

\usepackage{soul, url}

\newcommand{\pd}[2]{\frac{\partial #1}{\partial #2}}
\newcommand{\pdd}[2]{\frac{\partial^2 #1}{\partial  #2^2}}

\DeclareMathOperator{\trace}{trace}

\newcommand{\real}[1]{\mathbb{R}({#1})}
\newcommand{\imag}[1]{\mathbb{I}({#1})}

\newcommand{\bth}{{\bm \theta}}

\newcommand{\bte}{{\bm \epsilon}}

\newcommand{\alphaB}{\alpha_{\rm B}}

  \newcommand{\ws}{w_{\rm s}}
  \newcommand{\vs}{v_{\rm s}}
  \newcommand{\us}{u_{\rm s}}

  \newcommand{\V}{V}
  \newcommand{\Vr}{\V_{\rm ref}}

\newcommand{\uuf}{\mathbf{u}_{\rm f}}
\newcommand{\uue}{\mathbf{u}_{\rm e}}
\newcommand{\uus}{\mathbf{u}_{\rm s}}
\newcommand{\w}{\mathbf{w}}
\newcommand{\E}{\mathbf{E}}

\newcommand{\Tf}{\mathbf{T}_{\rm f}}
\newcommand{\pf}{p_{\rm f}}

\newcommand{\I}{\mathbf{I}}
\newcommand{\boldS}{\mathbf{S}}

\newcommand{\mufr}{\mu_{\mathrm{fr}}}
\newcommand{\mue}{\mu_{\mathrm{e}}}
\newcommand{\lambdae}{\lambda_{\mathrm{e}}}
\newcommand{\rhoe}{\rho_{\mathrm{e}}}

\newcommand{\kappafr}{\kappa_{\mathrm{fr}}}
\newcommand{\kappaf}{\kappa_{\mathrm{f}}}
\newcommand{\kappas}{\kappa_{\mathrm{s}}}
\newcommand{\rhoa}{\rho_{\mathrm{a}}}
\newcommand{\rhos}{\rho_{\mathrm{s}}}
\newcommand{\rhof}{\rho_{\mathrm{f}}}

\begin{document}

\title{Monitoring of water volume in a porous reservoir using seismic data: A 3D simulation study}

\author{M. Khalili${}^{a}$,   P. G\"oransson${}^{c}$, J.S. Hesthaven${}^{b}$,\\ A. Pasanen${}^{d}$, M. Vauhkonen${}^{a}$, and T. L\"ahivaara${}^{a}$\\
\normalsize ${}^{a}$Department of Technical Physics, University of Eastern Finland, Kuopio, Finland\\
\normalsize ${}^{b}$Computational Mathematics and Simulation Science,\\
\normalsize Ecole Polytechnique F\'ed\'erale de Lausanne, Lausanne, Switzerland\\
\normalsize ${}^{c}$Department of Aeronautical and Vehicle Engineering,\\
\normalsize KTH Royal Institute of Technology, Stockholm, Sweden\\
\normalsize ${}^{d}$Geological Survey of Finland, Kuopio, Finland
}

\maketitle

\section*{Abstract}

A potential framework to estimate the volume of water stored in a porous storage reservoir from seismic data is neural networks. In this study, the man-made groundwater reservoir is modeled as a coupled poroviscoelastic–viscoelastic medium, and the underlying wave propagation problem is solved using a three-dimensional discontinuous Galerkin method coupled with an Adams–Bashforth time stepping scheme. The wave problem solver is used to generate databases for the neural network-based machine learning model to estimate the water volume. In the numerical examples, we investigate a deconvolution-based approach to normalize the effect from the source wavelet in addition to the network's tolerance for noise levels. We also apply the SHapley Additive exPlanations method to obtain greater insight into which part of the input data contributes the most to the water volume estimation. The numerical results demonstrate the capacity of the fully connected neural network to estimate the amount of water stored in the porous storage reservoir.

\newpage

\section{Introduction}

Groundwater is one of the primary water sources for agriculture and domestic usage worldwide. However, the mismanagement and unsustainable use of existing water resources create diverse problems, such as water-level drawdown, deterioration of the water quality, and droughts. Therefore, the management of groundwater resources is a crucial component of sustainable development \cite{UN, sdg6}.

The characterization of groundwater storage using conventional methods, such as boreholes, coring, and trial pits (see, for example, \cite{prentice1990geology}), has always been challenging. These approaches are typically time consuming and expensive, and they provide spot data only (see, for example, \cite{baharuddin2018prediction, adedibu2019insight}). To remedy these challenges, a characterization using seismic data is commonly used to acquire subsurface hydrological parameters. As seismic waves propagate in the ground, they capture information about ground properties. This information includes, for example, reflections from the discontinuities and attenuation caused by the media. 

As a result, numerous geophysical studies have been devoted to predicting groundwater levels and groundwater storage. For example, a case study \cite{mcclymont2012locating} used a combination of geophysical techniques, such as ground-penetrating radar, seismic refraction, and time-lapse gravity surveying, to estimate groundwater storage in a field site in the Canadian Rockies. In \cite{clements2018tracking}, temporal and spatial variations in groundwater levels were predicted in San Gabriel Valley, California, using perturbations in seismic velocity inferred from ambient seismic noise. As a third example, we refer to \cite{mao2022space}, in which seismic interferometry techniques were used to infer the space–time evolution of relative changes in seismic velocity as a measure of hydrological parameters, including groundwater fluctuations. In this last example, the groundwater basins near Los Angeles, California, were used as the test areas. An extensive overview of the geophysical methods (in general) used for groundwater exploration and management is given in \cite{kirsch2006groundwater}. 

Groundwater exploration can be categorized into two types. One aims to search for new resources, and the other seeks to monitor existing water reservoirs. In this study, the main focus is on developing a methodology applicable for monitoring purposes. More speciﬁcally, the goal is to estimate the volume of water stored in a man-made, sandy aquifer using seismic full-waveform data. The case studied is motivated by a real site (sand infiltration in \cite{laukaa}), and based on prior knowledge, it is assumed that the material field is homogeneous and isotropic. The sand pool represents an unconﬁned water storage reservoir underlain by an impermeable bentonite lining. The source and receivers are placed on the ground surface for the studied three-dimensional (3D) benchmark problem. 

The water storage reservoir consists of two zones: the air-saturated and water-saturated zones. The water table (i.e., the surface between the air- and water-saturated zones) is expected to be sharp, with no partially saturated zone, and it is allowed to fluctuate freely. As a physical model for wave propagation in the saturated porous medium, Biot's isotropic poroviscoelastic model is used, while in a narrow zone next to the porous material, the isotropic viscoelastic solid model is employed. For a detailed discussion on the physics of wave propagation in porous and elastic media, please refer to \cite{love2013treatise, carcione15}.  
 
The approach studied in this work consists of two main components: forward and inverse problems. 
\begin{itemize}
    
\item The first component models wave propagation from the source to the receiver. In this study, wave propagation is modeled using the nodal discontinuous Galerkin (DG) method \cite{hesthaven_warburton_book}. The DG method is a well-suited approach to accurately simulate wave propagation. It can be used to accurately model material discontinuities, handle complex geometries, and enable efficient computations both in the central processing unit (CPU) and graphics processing unit (GPU) computational environments. The DG method has been studied in the context of wave problems by many authors (see, for example, \cite{delaPuente08_thesis}, \cite{WILCOX20109373}, \cite{zhan2019complete}, \cite{shukla2019}, \cite{XIE2019108865}, and \cite{ward2020discontinuous}). 

\item The inverse problem is solved using neural networks, which provide an efficient framework to analyze seismic data and have been examined by several authors. These studies include convolutional neural networks (CNNs) for earthquake detection and locations from seismograms \cite{perol}. Another example is a synthetic two-dimensional study \cite{lahivaara2019estimation}, in which CNNs were used to recover the water table level and amount of water stored in an aquifer. Finally, in \cite{Araya-Polo2020}, CNNs were used to estimate subsurface velocity profiles. For a recent review of neural networks on seismic data, please refer to \cite{mousavi}.
    
\end{itemize}
    
In this study, the use of neural networks is motivated by the fact that they can be trained to directly recover the amount of water stored from seismic data. Once the time-demanding training phase is done, the estimates can be obtained in almost real time even with a standard office computer, enabling the online monitoring of reservoir water  volume. 

Model interpretability is important to understand how different data components contribute to the estimates. For this purpose, we apply the SHapley Additive exPlanations framework \cite{lundberg2017unified} to investigate data contribution at the receiver level. Note that our intention with the SHAP analysis is to determine the contributions of different receivers rather than developing optimal receiver array configurations.

One of the common types of sources in seismic surveys is hitting a metal plate placed on the ground surface using a sledgehammer. This leads to an unknown source wavelet that, in turn, needs to be captured in the full waveform inversion. In principle, it is possible to record, for example, the vibrations of the steel plate, but these may contain data from multiple reflections and thus cannot be considered pure source wavelets anymore. It is also possible to consider the source wavelet as an additional unknown parameter (or parameters) that can be estimated during the inversion procedure. In this study, we use source-independent inversion using the deconvolution of the observed wavefield with a reference trace in the frequency domain, \cite{lee03, guo2019multi}. In this way, the effect of the shape of the source function can be eliminated from the traces.

\section{Problem description}\label{sec:prob}

\begin{figure*}[!htb]
\centering
\includegraphics[width=0.95\textwidth]{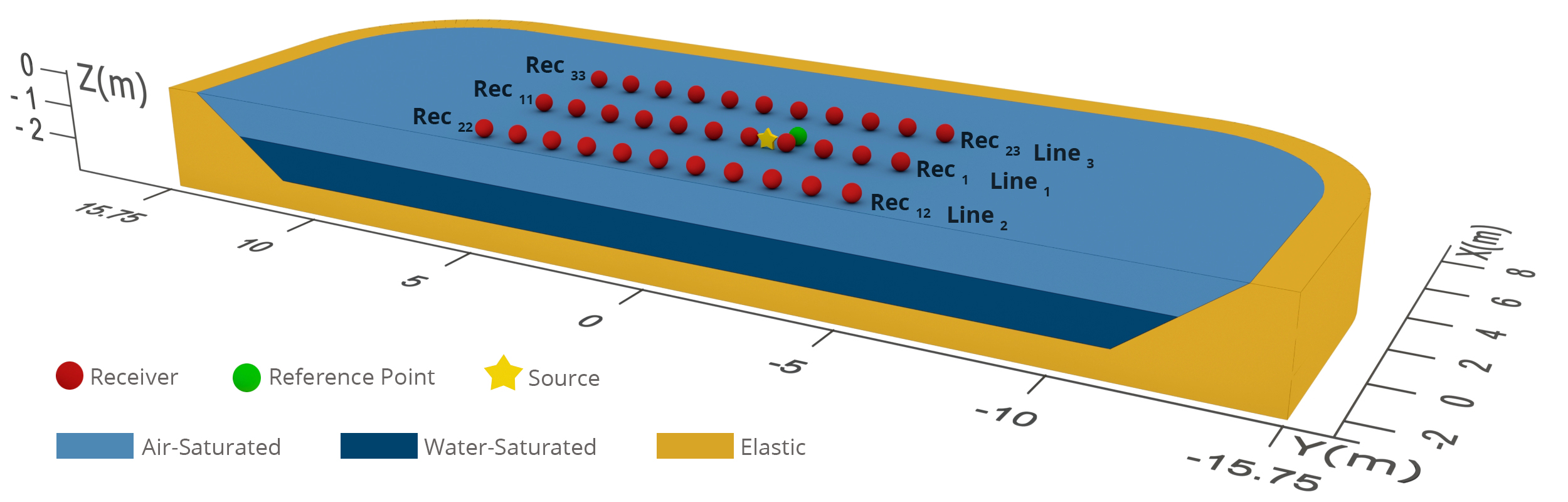}
\includegraphics[width=0.95\textwidth]{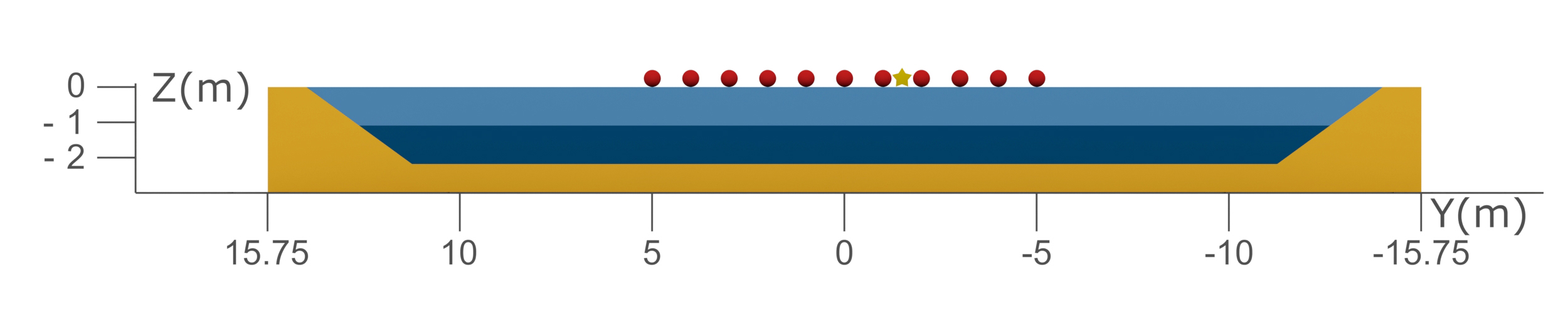}
\caption{A cross-section of the problem geometry. The light-blue color refers to the air-saturated zone, and the dark-blue color refers to the water-saturated zone. The light-brown color denotes the surrounding elastic material. The setup contains a total of 34 receivers, marked with red dots and one green dot, all of which are located on the ground surface. The source location is marked with a yellow star and placed between receivers 4 and 5. } \label{fig:graph}
\end{figure*}

As the problem geometry, a small-scale sand pool with known dimensions is studied. We also assume that the material is isotropic and homogeneous. Specifically, the geometry consists of a porous material and the surrounding elastic material (see Fig. \ref{fig:graph}). The porous material is divided into air- and water-saturated subdomains by a water table. The geometry studied mimics the sand infiltration area explained in \cite{laukaa}.

The geometry of the model is a  box with rounded corners. The dimensions are length 31.5 m, width 16.2 m, and height 2.75 m. The corners of the geometry are rounded, as shown in Fig. \ref{fig:graph}. The maximum length and width of the air-saturated zone are 29.5 m and 14.2 m, respectively. Finally, the bottom profile of the water-saturated zone is a rectangle with a length of 23.5 m and a width of 8.2 m. The bentonite-lined bottom is located at a depth of 2 m from the top surface. 

Figure \ref{fig:graph} shows the problem geometry with a water table level at $z = -1$ m. We assume a free boundary condition on the top surface, while other boundaries are modeled as absorbing boundaries. We also assume a total of 34 receivers to capture the solid velocity components $\vs$ and $\ws$ (i.e., the time derivatives of the corresponding solid displacements). Thirty-three of these receivers are divided into three lines, leading to 11 sensors on each line. Receiver lines are placed parallel to the $y$-axis. For each line, the $y$ coordinates are distributed uniformly over the range of $y\in \left[-5,\ 5\right]$ m, while the $x$-coordinate is 0.5 m for line 1, -2 m for line 2, and 3 m for line 3. One extra receiver, illustrated by a green dot symbol, is located at $(x, y)=(1, -2)$ m and is used as a reference point. A more detailed discussion of the reference point is given later in Section \ref{sec:matmod}. The source location is set to $(x,y)=(0.5, -1.5)$ m. The receivers and the source are placed on the ground surface ($z=0$ m). 

In this work, the seismic source is modeled as a point (impulse) source of force type. The force is directed along the negative $z$-axis. Two types of source wavelets are used, namely, a Gaussian 
\begin{equation}
\label{eq:gaussian}
    g = -\exp\left(a\left(t-t_0\right)^2\right),
\end{equation}
where $a = -(500\pi)^2$, and a first derivative of a Gaussian
\begin{equation}
\label{eq:1stderiv}
    g = \frac{(t-t_0)}{c}\exp\left(b\left(\left(t-t_0\right)^2-c^2\right)\right),
\end{equation}
where $b = -(f\pi)^2$ and $c=\sqrt{-0.5/b}$. The constant $c$ is used to normalize the source function amplitude to $[-1,1]$. In the following simulations, we set the frequency $f$ to 60 Hz, the time delay $t_0$ to $1.2/f$, and the modeling time from 0 s to 0.2 s.

\subsection{Physical parameters}\label{sec:params}

As the physical model for the wave problems investigated in this study, a coupled poroviscoelastic–viscoelastic model is used. The governing equations (i.e., Biot's poroviscoelastic wave equation and the viscoelastic wave equation) are given in Appendix \ref{sec:physnum}. 

The fluid parameters for the water-saturated subdomain are given by the density $\rhof = 1,000$ kg/m$^{3}$, the fluid bulk modulus $\kappaf = 2.1025$ GPa, and the viscosity $\eta = 1.14$e-3 Pa$\cdot$s, while in the air-saturated part, we set $\rhof = 1.2$ kg/m$^{3}$, $\kappaf = 1.3628$e5 Pa, and $\eta = 1.8$e-5 Pa$\cdot$s. The quality factor $Q_{\kappaf}$ is set to a large value. All other material parameters of the water storage reservoir are assumed to be randomized from uniform distributions. The minimum and maximum values are given in Table \ref{tab:vals}. Permeability is approximated from the Kozeny-Carman equation \cite{bear}
\begin{equation}
    k = \frac{D_p^2\phi^3}{180(1-\phi)^2},
\end{equation}
where the effective grain size $D_p$ of the sand is set to 0.0001 m. Finally, in the surrounding viscoelastic material, we assume that $\lambdae = 4.025$ GPa, $\mue = 1.6$ GPa, $\rhoe = 2500$ kg/m$^{3}$, $Q_P = 30$, and $Q_S = 30$. All material parameters are given for the reference frequency $\omega_r=120\pi$ Hz. Attenuation is modeled with $N=3$ mechanisms in both the water storage reservoir and the surrounding medium.

\begin{table}[!htb]
    \centering
    \caption{Material parameter bounds assumed for the water storage reservoir.}\label{tab:vals}
    \begin{tabular}{|c|c|c|c|c|c|}
    \hline
    variable name & symbol (unit)  & Minimum & Maximum\\
     &  & value & value\\
     \hline
    Solid density & $\rhos$ (kg/m${}^3$)& 2400 & 2800\\ 
     \hline
    Solid bulk modulus & $\kappas$ (GPa) & 45 & 55 \\
     \hline
     Frame bulk modulus  & $\kappafr$ (GPa) & 0.008 &  0.05 \\
     \hline
     Frame shear modulus   & $\mufr$ (GPa) & 0.002 & 0.04\\ 
     \hline
     Tortuosity   & $\tau$ & 1.1 & 1.6\\ 
     \hline
     Porosity   & $\phi$ ($\%$) &30 & 40\\
      \hline
      Quality factor   & $Q_{\mufr}$ &15 & 50\\
      \hline
      Quality factor   & $Q_{\kappas}$ &80 & 120\\
      \hline
      Quality factor   & $Q_{\kappafr}$ &15 & 50\\
      \hline
    \end{tabular}
    \label{tab:numbers}
\end{table}

The water volume in the porous storage reservoir can be calculated by multiplying the volume of the water-saturated domain by the porosity. In this study, the amount of water is calculated only from the water-saturated zone located exactly under the array or receivers. 

\subsection{Source function normalization}\label{sec:matmod}

The observation model 
\begin{eqnarray}
    \V& = &\mathcal{A}(\boldsymbol{m}) + e\\
\label{eq:obsmod}
     &=& X + e,
\end{eqnarray}
consists of the measurement data vector $\V=[\vs, \ws]^\top$, the forward model $\mathcal{A}$, the model parameters $\boldsymbol{m}$, and the noise $e$. The noise model is assumed to be Gaussian (see Section \ref{sec:trval} for more details).

For the applied inversion technique, the seismic source function needs to be known. In this study, the effect from the source wavelet is normalized via a deconvolution operation. For this operation, transient signals on each receiver are first transformed to the frequency domain. As a system response function for the deconvolution, data from the additional reference data point, as shown in Fig. \ref{fig:graph}, are used. Thus, the above observation model (\ref{eq:obsmod}) is replaced by
\begin{eqnarray}
    \frac{F(\V)}{F(\Vr)} &=& \frac{F(X)}{F(X_{{\rm ref}})} + \hat{e}\\
    \label{eq:obf}
\hat{V} &=& \hat{X} + \hat{e},
\end{eqnarray}
where $F$ denotes the Fourier transform, subscript $\rm{ref}$ denotes the data on the reference point, and $\hat{e}$ denotes the noise in the frequency domain formulation after the normalization. Equation (\ref{eq:obf}) may become ill conditioned when the denominator approaches zero. Thus, as in \cite{wen}, Wiener filtering is used to regularize the model. For the regularization coefficient, we use 0.001.

\section{Training and validation databases and noise model}\label{sec:trval}

We use a nodal discontinuous Galerkin (DG) method coupled with third-order Adams–Bashforth time stepping as the numerical method to approximate wave propagation. More detailed discussion of the numerical method and implementation details are given in Appendix \ref{sec:dg}. 

The initial step of the (supervised) machine learning (ML) algorithm implementation is the creation of the training and validation databases. The generation procedure of physical parameters is used to collect each sample in the training, validation, and test datasets. The acquisition geometry of forward modeling on each sample is the same as that described previously. We create 15,000 training samples using computational grids with approximately 2 elements per shortest, inviscid material model-based wavelength. For each element, fifth-order polynomial basis function orders are used. An extra validation database of 3,000 samples is also prepared to assess the models during the training phase and monitor the network's generalization capabilities. Similarly, as with the training samples, we use fifth-order polynomials with the DG solver, but the element size criterion is relaxed to approximately 1.9 elements per shortest wavelength. We use the Gaussian source function (\ref{eq:gaussian}) for all training and validation samples. 

All samples of the train and validation databases are contaminated with Gaussian noise. To be specific, we make five copies of each sample in the database and add noise for a sample as follows:
\begin{equation}
\label{eq:noisemodel}  
X_{\ell}^\textrm{noised}=X_\ell+A\alpha\epsilon^A+B|X_{\ell}|\epsilon^B,
\end{equation}
where $\epsilon^A\sim\mathcal{N}(0, 1)$ and $\epsilon^B\sim\mathcal{N}(0, 1)$ are Gaussian random vectors, and $\alpha$ is the largest absolute value of the training database. The second term represents additive (stationary) white noise. The coefficients $A$ and $B$ are randomized from the uniform distribution for each sample. To include a wide range of noise levels (NLs) that are compatible with the amplitudes of the real noises, the standard deviation of the white noise component varies uniformly in $[0.1, 0.5]\%\alpha$, and the standard deviation of the relative component is from $[0,  2]\% |X_\ell|$. The total number of samples in the training database is $5\times 15,000 = 75,000$, and in the validation database, it is $5\times 3,000 = 15,000$.

\section{Neural networks}

In this study, we use a fully connected neural network to recover the amount of water stored (i.e., one output value) from the synthetic seismic data. The simulated data $\V$ are down-sampled to a sampling frequency of 4 kHz, contaminated by noise, and then transformed to the frequency domain and deconvoluted with the reference data. The input data for the neural network algorithm contain the stacked real and imaginary parts (i.e. $[\real{\hat{X}}, \imag{\hat{X}}]^\top$ or $[\real{\hat{V}}, \imag{\hat{V}}]^\top$). From the frequency spectrum, we select 12 frequencies varying from 35 Hz to 90 Hz, leading to a total input size of $33\times12\times2\times2 = 1,584$. 

An important part of an ML system is the choices for the network architecture. In this study, the architecture is selected as follows. The optimizer algorithm is set to that in Adam \cite{kingma14}. We assume that the architecture contains either one, two, three, or four hidden layers. For each layer choice, we run KerasTuner \cite{omalley2019kerastuner} with Bayesian optimization search algorithms to optimize the activation function, the learning rate, and the number of neurons per hidden layer. As the possible activation function, we allow KerasTuner to choose "relu," "sigmoid," "tanh," "selu," "swish," "exponential," or "LeakyReLU." The number of neurons per layer is allowed to vary from 4 to 1,504, and for the learning rate, we let KerasTuner choose [1e-3, 1e-4, 1e-5, 1e-6]. In addition to the previous steps, for each layer configuration, we run KerasTuner for four batch size values [32, 64, 128, 256]. The best-performing network (mean absolute error (MAE) with validation) is used as a selection criterion for the architecture. 

After the hyperparameter tuning, the final architecture consists of three hidden layers with 1,034 (layer 1), 594 (layer 2), and 184 (layer 3) neurons, a swish activation function, a learning rate of 1e-4, and a batch size of 128. For the output layer, we use linear activation. Early stopping is also activated both in the hyperparameter tuning and actual training phases. We use the Python library TensorFlow \cite{tensorflow2015-whitepaper} as a computing interface with Keras backend \cite{chollet2015keras} for programming.

\section{Results}

As a test dataset for the trained neural network, an additional 3,000 samples are used. To ensure different numerical noises in the database, we use sixth-order basis orders for the DG solver, in addition to 2.2 elements per shortest (inviscid-based) wavelength criterion for grid density. We also assume a first-derivative Gaussian function (\ref{eq:1stderiv}) as a source wavelet. 

\subsection{Wave fields and seismograms}

\begin{figure*}[!htb]
\includegraphics[width=0.95\textwidth]{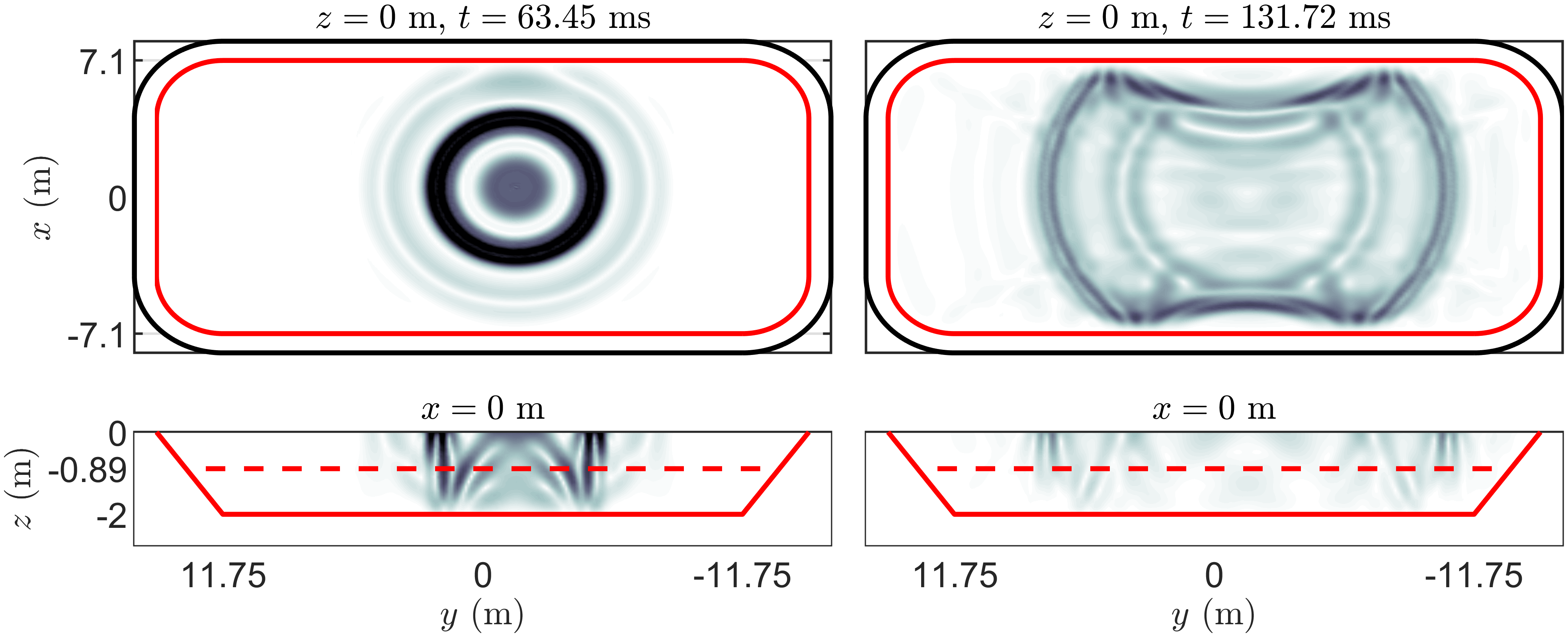}
\centering
\caption{Snapshots of the total velocity field on two cross-sections at two time instants. The zero coordinate axis and time instant are shown in the title. In the figures, the solid red line denotes the surface between the water storage reservoir and the surrounding material, and the dashed red line is the water table level ($z = -0.89$ m). On the top row, the solid black line shows the boundary for the exterior surface. } \label{fig:snap}
\end{figure*}

The snapshots of a noiseless total velocity field $\sqrt{\us^2+\vs^2+\ws^2}$ from one sample in the test database are shown in Fig. \ref{fig:snap}. For the velocity field, $\us$ is the $x$-directional component of the solid velocity. For the visualized case, the water table level is at $z = -0.89$ m. The snapshots are shown at two time instants: 63.45 ms (left panel) and 131.72 ms (right panel). At the first time instant, incident waves have reached the surface between the air-saturated zone and the surrounding elastic material at $x = 7.1$ m. We also observe the first reflection on the ground surface that originates from the bottom surface of the water storage. Snapshots on the right panel show the strong reflection from the surface between the air-saturated zone and the surrounding elastic material. The effect from the water table level can be seen more clearly in the $x=0$ cross-section graph. 

The test data comprise four different NLs to study the network's sensitivity to noise contamination. The same noise model (\ref{eq:noisemodel}), as used with the training and validation samples, is also used with the test samples, but the parameters $A$ and $B$ are fixed instead of randomized. The parameter choices are shown in Table \ref{tab:vals_noise}. Combinations are selected in such a way that the first NL corresponds to the very small noise amplitude, the second combination leads to a moderate NL, the third to a high noise amplitude, and the fourth to an extreme NL. The values of the fourth NL exceed the values used for the training database.
\begin{table}[!htb]
    \centering
    \caption{Noise amplitude parameters $A$ and $B$ for different noise levels (NLs).}\label{tab:vals_noise}
    \begin{tabular}{|c|c|c|c|c|c|}
    \hline
    Noise parameter & NL$_1$  & NL$_2$ & NL$_3$ & NL$_4$ \\
     \hline
    $A$ & 0.001 & 0.0025 & 0.005 & 0.01\\ 
     \hline
    $B$ & 0.0 & 0.01 & 0.02 & 0.05\\
      \hline
    \end{tabular}
\end{table}

\begin{figure*}[!htb]
\includegraphics[width=0.95\textwidth]{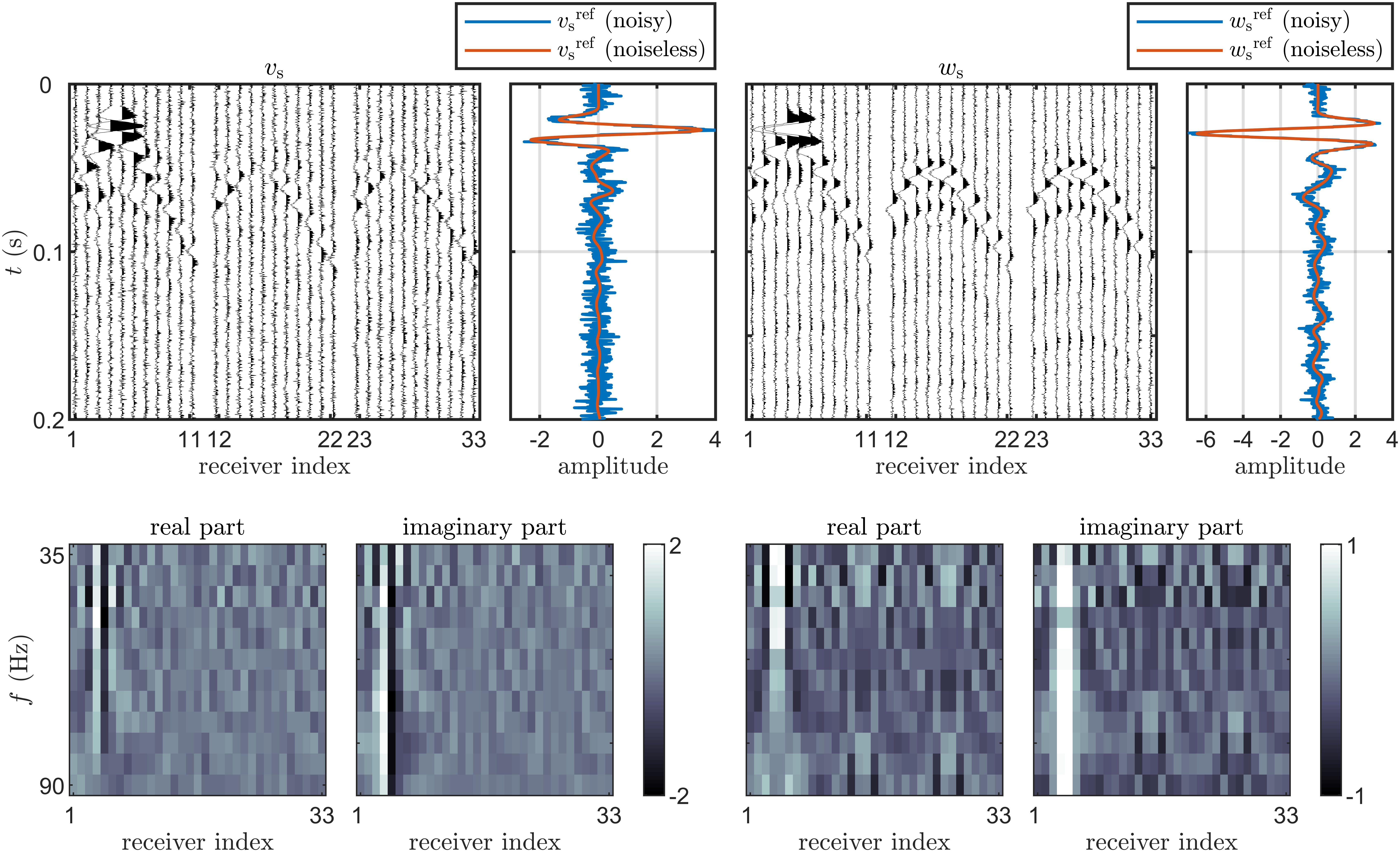}
\centering
\caption{An example sample in the test database. Transient data are shown on the top row, and frequency domain data are shown on the bottom row. The left column shows the $\vs$ component, and the right column shows the $\ws$ component.} \label{fig:signals}
\end{figure*}

Figure \ref{fig:signals} shows the noisy seismograms for the same model, as depicted in Fig. \ref{fig:snap}. As for the NL, we assume NL$_3$. The velocity components on the reference point are shown both with and without noise to illustrate the NL. The bottom row shows the data in the frequency domain, separately for the real and imaginary parts of the velocity components.

\subsection{Predictions of water volume}

\begin{figure*}[!htb]
\includegraphics[width=0.95\textwidth]{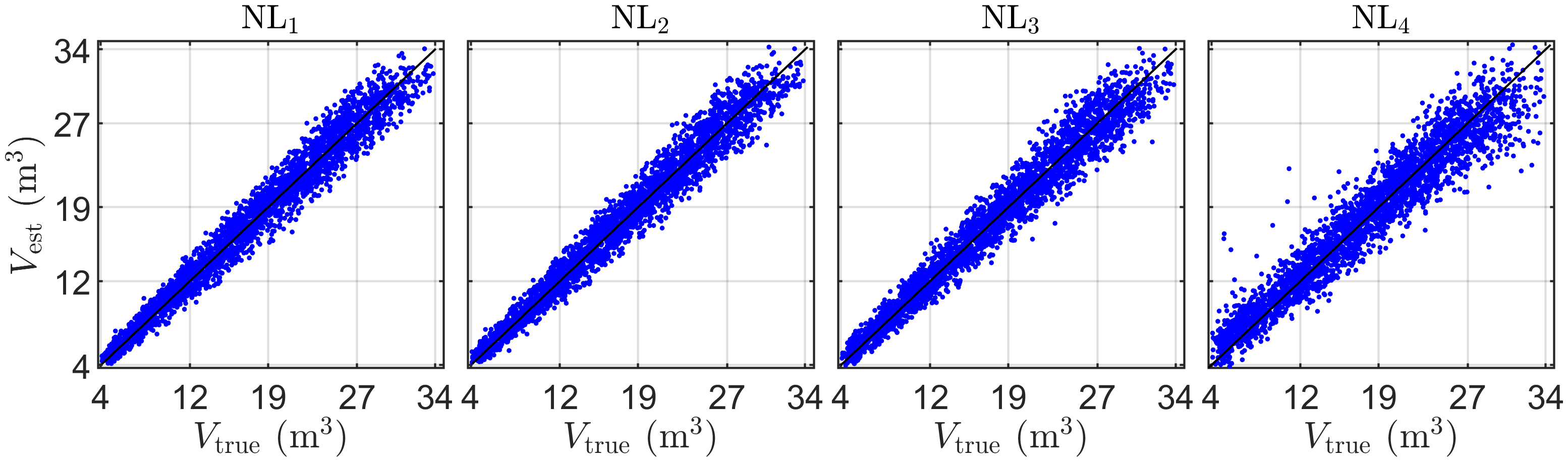}
\includegraphics[width=0.95\textwidth]{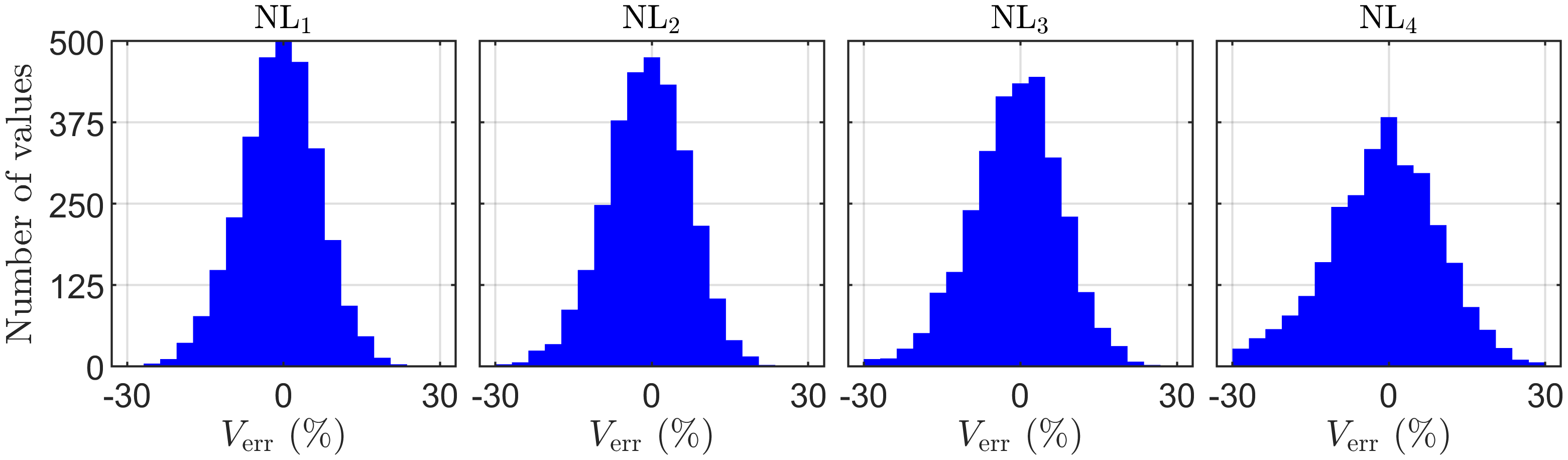}
\centering
\caption{Estimated volumes of water as functions of true value and the relative error histogram for four NL configurations. The NL varies from low to extreme from left to right.} 
\label{fig:results_n1}
\end{figure*}

Figure \ref{fig:results_n1} shows the true water volume $V_{\rm{true}}$ (cubic meters) versus its predicted values $V_{\rm{est}}$ (cubic meters), with each figure accompanied by a histogram of relative prediction errors $V_{\rm{err}}$. The test dataset has four levels of noise numbered in Table \ref{tab:vals_noise} corresponding to the respective panels in Fig. \ref{fig:results_n1}. With increasing noise, the histograms slightly spread, which is concretely shown in Table \ref{tab:bias} with the MAE and root mean square error (RMSE) as criteria. Clearly, this can be seen in the case of the extreme noise amplitude NL$_4$, which goes beyond the values used in the training stage.

\subsubsection*{SHAP analysis}

\begin{figure*}[!htb]
\includegraphics[width=0.95\textwidth]{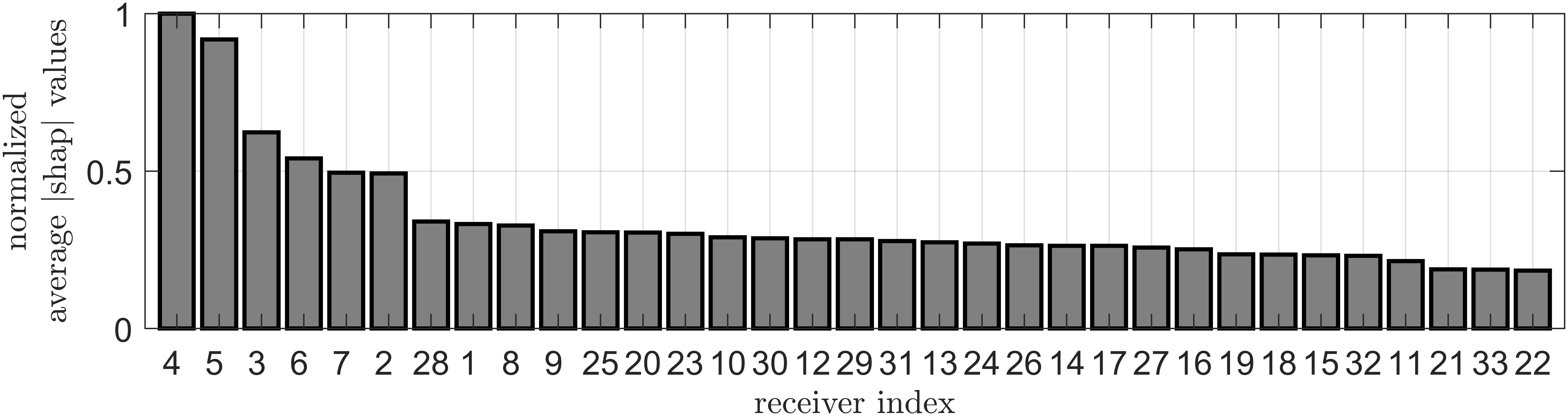}
\centering
\caption{Average of the absolute SHAP values for each receiver. Values are sorted such that the most contributing receivers are on the left.} \label{fig:shap}
\end{figure*}

To investigate the contribution of individual data inputs, we use a feature contribution method called SHAP to explain the predictions. We use the Python library called Shap \cite{lundberg2017unified} and apply the DeepExplainer algorithm for 10,000 randomly selected samples from the training database to build the explainer model. The model is then applied to a randomly selected set of 1,000 samples from the test database (NL$_2$). The model gives the SHAP value for each unit of the input data (see Fig. \ref{fig:signals} (bottom)), but here, we are more interested in investigating the data contribution at the receiver level. Therefore, we take the absolute value of each SHAP value and calculate the average for each receiver over the test samples, as shown in Fig. \ref{fig:shap}. The results show that receivers 4, 5, 6, 3, 7, and 2 contribute the most to predicting the model's output. 

Next, we construct two new receiver configurations and train the neural network model for these. For the first configuration, we select six receivers having the largest SHAP values (see Fig. \ref{fig:shap}), and for the second configuration, we randomly select six receivers from the full sensor array. The network architecture consists of three hidden layers, a learning rate of 1e-4, a swish activation function, and a batch size of 128, as used with the full receiver array. The number of neurons per hidden layer was optimized with KerasTuner. The network with SHAP analysis-based receiver selection consists of 1,394 (layer 1), 894 (layer 2), and 564 (layer 3) neurons and with randomly selected receivers of 1,394 (layer 1), 1,064 (layer 2), and 824 (layer 3) neurons. Note that we have used the same training database as with the full receiver array, just selecting the corresponding limited receiver array indices.

\begin{figure*}[!htb]
\includegraphics[width=0.95\textwidth]{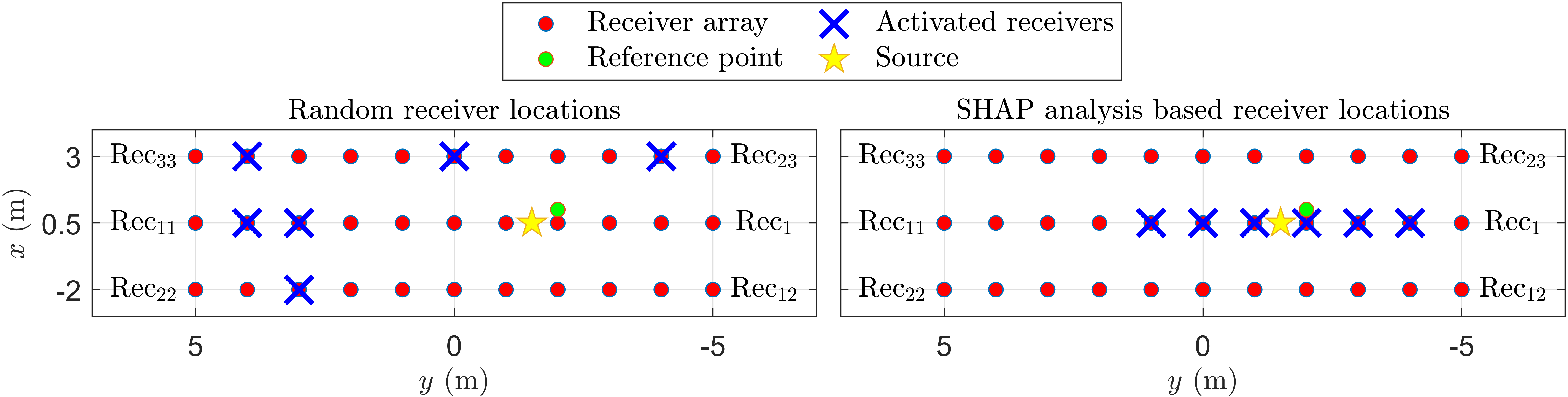}
\includegraphics[width=0.95\textwidth]{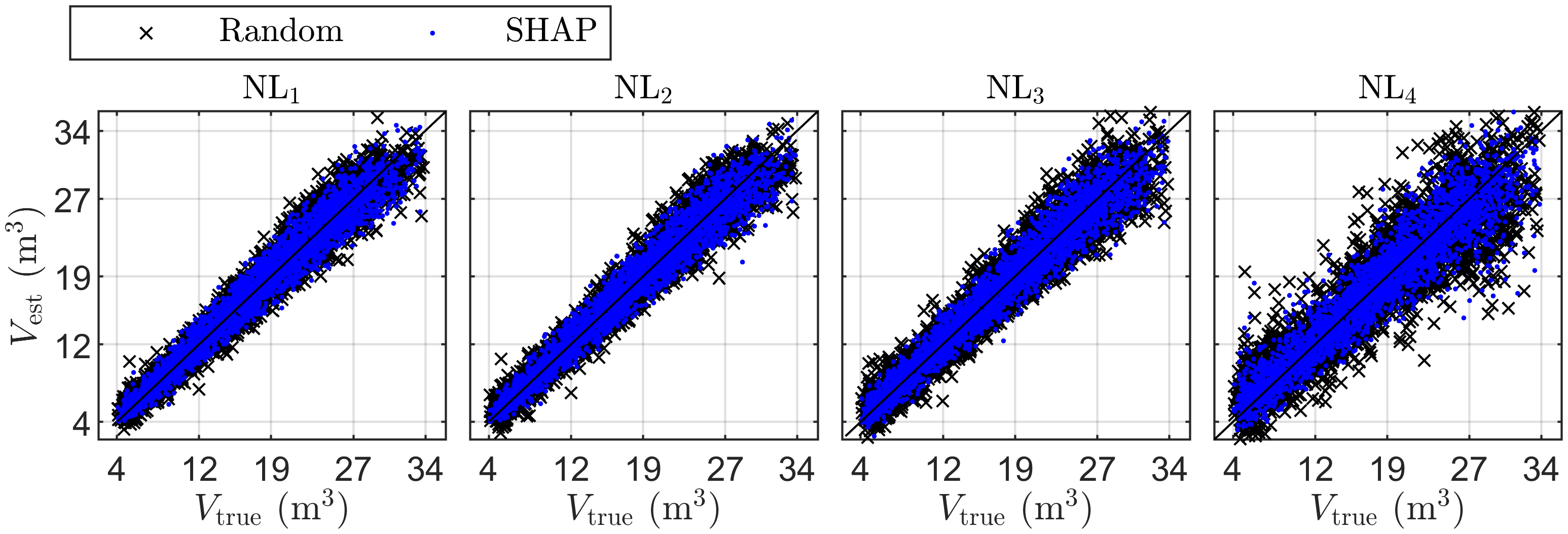}
\centering
\caption{TopLeft: Randomly selected and TopRight: Six SHAP value-based receivers. Estimated volumes of water as a function of the true value for four NL configurations. In the bottom row, the NL varies from low to extreme from left to right.}\label{fig:results_n1_shap}
\end{figure*}

Figure \ref{fig:results_n1_shap} shows the estimated water  volume as a function of the true water volume for both receiver configurations. The results indicate that the estimation accuracy is improved when using the SHAP analysis-based receivers. We can also see that the randomly selected receiver configuration leads to more outlier values. 

Table \ref{tab:bias} lists the bias, MAE, and RMSE values for all three receiver configurations and the four NLs. From the table, we can see that the full receiver array leads to the most accurate results.  
To be more specific, the average values of the RMSE for full, SHAP, and random are 1.5, 1.8, and 2.0, respectively. Finally, the noise tolerance is slightly worse with the randomly selected receivers.

\begin{table}[!htb]
    \centering
    \caption{The biases, mean absolute errors (MAEs), and root mean square errors (RMSEs) for different nose levels (NLs) were evaluated with the test database. The first four lines are for the full receiver array, the following four lines are for the most important SHAP analysis-based six receivers, and the last four lines are for the six randomly selected  receivers.}\label{tab:bias}
    \begin{tabular}{|l|c|c|c|c|c|}
    \hline
   Receiver  & Noise level & Bias  & MAE & RMSE \\
   array setup & & (cubic meters) & (cubic meters)  & \\
     \hline
     \parbox[t]{2mm}{\multirow{4}{*}{full}} & NL$_1$ & -0.0760  &  0.9970   & 1.2915 \\ 
     \cline{2-5}
    & NL$_2$ & -0.0876  &  1.0312   & 1.3286 \\
      \cline{2-5}
   & NL$_3$ & -0.0525  &  1.1321  &  1.4605 \\
      \cline{2-5}
   & NL$_4$ & -0.0807  &  1.4958  &  1.9728 \\\hline\hline
  \parbox[t]{2mm}{\multirow{4}{*}{SHAP}} & NL$_1$ &  -0.0290  &  1.1481  &  1.4933\\
  \cline{2-5}
   & NL$_2$ & -0.0248  &  1.1980  &  1.5698\\\cline{2-5}
   & NL$_3$ & -0.0421  &  1.3059  &  1.7087\\\cline{2-5}
   & NL$_4$ &  0.0373  &  1.7965  &  2.4338\\\hline\hline
  \parbox[t]{2mm}{\multirow{4}{*}{random}} & NL$_1$ &  -0.0776  &  1.1935 &  1.5382\\
  \cline{2-5}
   & NL$_2$ & -0.0636  &  1.2542  &  1.6226\\\cline{2-5}
   & NL$_3$ & 0.0114 &   1.4780 &   1.9232\\\cline{2-5}
   & NL$_4$ & 0.0935  &  2.2569 &   3.0251\\\hline
    \end{tabular}
\end{table}

\section{Conclusions}

This study estimated the volume of water stored in a 3D porous domain from synthetic seismic data. The study was purely numerical, while the problem geometry was motivated by a real test site. As a physical model for wave propagation, a coupled poroviscoelastic–viscoelastic model, which was solved with the nodal DG method, was used. The applied software simulated wave propagation in the time domain, but in this study, the data were transformed to the frequency domain, enabling us to normalize the effect from the source wavelet. The fully connected neural networks were used to recover the water volume values from the frequency domain data. 

The problem was formulated so that knowledge about the actual source function is not needed. Knowing the frequency content of the test data is sufficient to build the training data of the neural network with the corresponding frequency band. In the studied deconvolution-based approach, the shape of the source function was not used when solving the inverse problem because the source function was normalized in the same way for all data samples.

The results of this study support the use of a fully connected network to estimate the water storage reservoir parameters of interest with a wide selection of noise amplitudes, while inconvenient parameters can profitably be marginalized. Compared with traditional full waveform inversion, which requires the estimation of geometry and porosity, the suggested approach allows water volume  estimation with a significant reduction in computational and labor costs. Nevertheless, one must note here that the building of a training database for the studied (supervised) neural networks requires much computational time and resources, but once it is done, and the network is trained, the estimates can be computed in almost real time without powerful computational resources. 

With the current problem setup, in principle, we can estimate the water volume accurately in the entire water-saturated subdomain, but this cannot be generally assumed in more realistic aquifer scenarios. This is related to the specific feature of the setup studied, in which porous water storage is fully surrounded by an elastic domain that strongly reflects waves back to the porous domain. 

We performed SHAP analysis on the developed ML model to determine which receivers contributed the most to estimating the stored water volume. We also tested the effect of selecting much fewer receivers than those used in the full receiver array. Here, the selection was made based on the SHAP analysis and then randomly. It was evident that for the current problem setup, the characterization method, together with the selection of the receivers, was quite robust. We also wanted to emphasize that the SHAP analysis would most likely lead to totally different results when the problem geometry is more complex and if the material fields are heterogeneous. 

The ability of neural networks to extrapolate beyond training data is often very poor. In the context of this study it is assumed that if the geometry used to create the test data deviates significantly from the geometry used for the training data, the estimates produced by the neural network will be biased. Similarly, if the physical parameters are significantly inhomogeneous or deviate far from the a priori model of the training material, bias may also occur.  One possible research direction for the future is to investigate methods that could be used to assess the reliability of these estimates.

Future research could include a study with real data, their uncertainties, and a receiver-level SHAP analysis. It is possible that for real data, we may need to, for example, update the prior model for the material, use a more complicated noise model, and consider different network architectures, but these are topics for future research.

\appendices

\section{Governing equations of the physical models}\label{sec:physnum}

\subsection{Biot's poroelastic wave equation with attenuation}\label{sec:biot}

In this study, we use  Biot's wave equation \cite{carcione15}
\begin{eqnarray}
  \label{eq:biot2a}
  \rhoa \pdd{\uus}{t} +\rhof \pdd{\w}{t} &=&\nabla\cdot {\bf T},\\
\label{eq:biot2b}
\rhof \pdd{\uus}{t} + m\pdd{\w}{t} +  \frac{\eta}{k}\pd{\w}{t}
  &=&\nabla\cdot \Tf,  
\end{eqnarray}
to model wave propagation in 3D poroelastic media. In model (\ref{eq:biot2a}) and (\ref{eq:biot2b}), $\uus$ is the solid displacement, $\uuf$ is the fluid displacement, and $\w$ is the relative displacement of fluid $\w=\phi(\uuf -\uus )$, where $\phi$ is the porosity. In addition, $\rhos$ is the solid density, $\rhof$ is the fluid density, $\rhoa = (1-\phi)\rhos + \phi \rhof$ is the average density, $\eta$ is the viscosity, and $k$ is the permeability. In (\ref{eq:biot2b}), $m = \rhof \tau/\phi$, where $\rhof$ is the fluid density and $\tau$ is the tortuosity.

The lossless solid and fluid stress tensors ${\bf T}$ and $\Tf$ 
\begin{equation}
  {\bf T}=
  \begin{pmatrix}
    T_{11} &T_{12} & T_{13}\\
    T_{12} &T_{22} & T_{23}\\
    T_{13} &T_{23} & T_{33}\\
  \end{pmatrix}, \quad   {\Tf}=
  \begin{pmatrix}
    -\pf & 0 & 0\\
    0 & -\pf & 0\\
    0 &0 & -\pf\\
  \end{pmatrix}, 
\end{equation}
and can be expressed as
\begin{eqnarray}
\label{eq:T}
  {\bf T} &=& 2\mufr {\bf E} + \left(\lambda\trace({\bf E}) 
  - \alphaB M\zeta\right) \I_3,\\
\label{eq:Tf}
  \Tf &=& \left(\alphaB M\trace({\bf E}) - M\zeta\right) \I_3,
\end{eqnarray}
where $\mufr$ is the frame shear modulus, $\zeta=-\nabla\cdot \w$ is the variation in fluid content, and $\I_{3}$ denotes an identity matrix of size $3\times 3$. In addition, $\lambda$ denotes a Lam\'e parameter and can be written as
\begin{equation}
\label{eq:lambda}
\lambda = \kappafr + \alphaB^2 M -\frac{2}{3}\mufr,
\end{equation}
where $\kappafr$ denotes the frame bulk modulus. Let $\kappas$ denote the solid bulk modulus, where we can write
\begin{equation}
\label{eq:alpha}
\alphaB = 1 - \frac{\kappafr}{\kappas}
\end{equation}
and
\begin{equation}
\label{eq:bigM}
  M =\frac{\kappas}
  {\alphaB- \phi(1-\kappas/\kappaf)}, 
\end{equation}
where $\kappaf$ is the fluid bulk modulus. In Eqs. (\ref{eq:T}) to (\ref{eq:Tf}), $\E$ denotes the solid strain tensor and can be written as
\begin{equation}
  \E =
  \begin{pmatrix}
    \epsilon_{11} &\epsilon_{12} & \epsilon_{13}\\
    \epsilon_{12} &\epsilon_{22} & \epsilon_{23}\\
    \epsilon_{13} &\epsilon_{23} & \epsilon_{33}\\
  \end{pmatrix},
\end{equation}
where $\epsilon_{ij},\ i,j=1,2,3$ are the components of the tensor (for a detailed derivation, see \cite{carcione15}).

As discussed in \cite{carcione15}, viscoelasticity can be introduced into Biot's poroelastic equations to model attenuation mechanisms related to kinetic energy (viscodynamic dissipation) and potential energy (stiffness dissipation). 

The viscodynamic dissipation effect can be further divided into the low- and high-frequency regimes. The main difference between these regimes is that for low frequencies, the Pouiselle assumption is valid, and inertial forces are negligible to viscous forces. At higher frequencies, inertial forces are no longer negligible, and the viscous resistance to fluid flow given by the coefficient of the dissipative term is frequency dependent. Biot's characteristic frequency of $f_c$ can be used to distinguish between these frequency regimes:
\begin{equation}
\label{eq:regime}
  f_c = \frac{\eta \phi}{2 \pi \tau \rhof k}.
\end{equation}

The stiffness dissipation and the applied model are discussed next in Section \ref{sec:stiffdyn}.

\subsubsection{Stiffness dissipation}\label{sec:stiffdyn}

In this work, stiffness attenuation is modelled using the generalized Maxwell body (GMB) rheology. For a detailed discussion of the GMB, please refer to \cite{moczo2014finite}. To include stiffness dissipation in the poroelastic media, we follow \cite{zhan2019complete} and re-write solid tensor ${\bf T}$ components as
\begin{align}
\label{eq:e11_mod}
  T_{11} &= \left(\lambda + 2\mufr\right)^U \epsilon_{11} + \lambda^U \epsilon_{22} + \lambda^U \epsilon_{33} - (\alphaB M)^U\zeta - \nonumber \\
  &  
  \sum_{n=1}^N\left(\left(\lambda + 2\mufr\right)^U Y_{\lambda+2\mufr}^{(n)}\theta_{\epsilon_{11}}^{(n)} + \lambda^U Y_{\lambda}^{(n)}\theta_{\epsilon_{22}}^{(n)} + \lambda^U Y_{\lambda}^{(n)}\theta_{\epsilon_{33}}^{(n)} - (\alphaB M)^UY_{\alphaB M}^{(n)}\theta_{\zeta}^{(n)}\right),\\
  T_{22} &=  \lambda^U \epsilon_{11} + \left(\lambda + 2\mufr\right)^U\epsilon_{22} + \lambda^U \epsilon_{33} - (\alphaB M)^U\zeta - \nonumber \\
  &  
  \sum_{n=1}^N\left(\lambda^U Y_{\lambda}^{(n)}\theta_{\epsilon_{11}}^{(n)} + \left(\lambda + 2\mufr\right)^U Y_{\lambda+2\mufr}^{(n)}\theta_{\epsilon_{22}}^{(n)} + \lambda^U Y_{\lambda}^{(n)}\theta_{\epsilon_{33}}^{(n)} - (\alphaB M)^UY_{\alphaB M}^{(n)}\theta_{\zeta}^{(n)}\right),
\\
  T_{33} &=  \lambda^U \epsilon_{11} + \lambda^U\epsilon_{22} + \left(\lambda + 2\mufr\right)^U \epsilon_{33} - (\alphaB M)^U\zeta - \nonumber \\
  &  
  \sum_{n=1}^N\left(\lambda^U Y_{\lambda}^{(n)}\theta_{\epsilon_{11}}^{(n)} + \lambda^U Y_{\lambda}^{(n)}\theta_{\epsilon_{22}}^{(n)} + \left(\lambda + 2\mufr\right)^U Y_{\lambda+2\mufr}^{(n)}\theta_{\epsilon_{33}}^{(n)} - (\alphaB M)^UY_{\alphaB M}^{(n)}\theta_{\zeta}^{(n)}\right),\\
  T_{12} &= 2\mufr^U \epsilon_{12} - 
  \sum_{n=1}^N2\mufr^U Y_{\mufr}^{(n)}\theta_{\epsilon_{12}}^{(n)},\\
  T_{13} &= 2\mufr^U \epsilon_{13} - 
  \sum_{n=1}^N2\mufr^U Y_{\mufr}^{(n)}\theta_{\epsilon_{13}}^{(n)},\\
  T_{23} &= 2\mufr^U \epsilon_{23} - 
  \sum_{n=1}^N2\mufr^U Y_{\mufr}^{(n)}\theta_{\epsilon_{23}}^{(n)},  
  \end{align}
and $\Tf$ components as   
\begin{align}
\label{eq:Te11_mod}
-\pf &= \left(\alphaB M\right)^U \epsilon_{11} + \left(\alphaB M\right)^U \epsilon_{22} + \left(\alphaB M\right)^U \epsilon_{33} - M^U\zeta\nonumber \\ 
& - \sum_{n=1}^N\left(\left(\alphaB M\right)^U Y_{\alphaB M}^{(n)}\theta_{\epsilon_{11}}^{(n)} + \left(\alphaB M\right)^U Y_{\alphaB M}^{(n)}\theta_{\epsilon_{22}}^{(n)} + \left(\alphaB M\right)^U Y_{\alphaB M}^{(n)}\theta_{\epsilon_{33}}^{(n)} - M^U Y_{M}^{(n)}\theta_{\zeta}^{(n)}\right).
\end{align}
In (\ref{eq:e11_mod})–(\ref{eq:Te11_mod}), the superscript $U$ denotes an un-relaxed material parameter. The formulas for obtaining un-relaxed material parameters are given later in Section \ref{sec:unrelaxed}.

The $7N$ memory variables $\bth^{(n)} = (\theta_{\epsilon_{11}}^{(n)}, \theta_{\epsilon_{22}}^{(n)}, \theta_{\epsilon_{33}}^{(n)}, \theta_{\epsilon_{12}}^{(n)}, \theta_{\epsilon_{13}}^{(n)}, \theta_{\epsilon_{23}}^{(n)}, \theta_{\zeta}^{(n)})$ introduced in (\ref{eq:e11_mod})-(\ref{eq:Te11_mod}) satisfy the ordinary differential equations
\begin{equation}
\label{eq:thetaeqs}
\frac{\partial \bth_{j}^{(n)}}{\partial t} = \omega^{(n)}\left(\bte_j-\bth_{j}^{(n)}\right),\quad j = 1,\ldots,7,
\end{equation}
where $\omega^{(n)}$ are the relaxation frequencies of the different mechanisms, and\\
$\bte = (\epsilon_{11}, \epsilon_{22}, \epsilon_{33}, \epsilon_{12}, \epsilon_{13}, \epsilon_{23}, \zeta)$.

\subsubsection{$Q$ Transformation rule}

In the stiffness dissipation model, we assume that the quality factors for the attenuation of the shear $Q_{\mufr}$ and bulk $Q_{\kappafr}$ moduli in the frame, the solid–fluid coupling $Q_{\kappas}$, and the bulk modulus of the saturated fluid $Q_{\kappaf}$ are known. We denote $Q = \left(Q_{\mufr}, Q_{\kappas}, Q_{\kappafr}, Q_{\kappaf}\right)$ and follow the transformation rule steps given in \cite{zhan2019complete}. 

The Kjartansson model \cite{Kjartansson} enables us to calculate the complex-valued, frequency-dependent parameters $\bar{M}(\omega) = \left(\bar{M}_{\mufr}, \bar{M}_{\kappas}, \bar{M}_{\kappafr}, \bar{M}_{\kappaf}\right)$, which can be written as 
\begin{eqnarray}
\label{eq:kjar}
\bar{M}_j = M^R_j\cos^2\left(\frac{\pi\gamma}{2}\right)\left(\frac{i\omega}{\omega_r}\right)^{2\gamma},\quad j=1,\ldots,4,
\end{eqnarray}
where $\gamma = \pi^{-1}\arctan\left(Q_j^{-1}\right)$. In (\ref{eq:kjar}), $M^R= \left(M_{\mufr^R}, M_{\kappas^R}, M_{\kappafr^R}, M_{\kappaf^R}\right)$ denote vectors containing the material parameters obtained at a reference frequency of $\omega_r$. Next, the complex-valued parameters are substituted to (\ref{eq:lambda}), (\ref{eq:alpha}), and (\ref{eq:bigM}) to obtain the complex-valued $(\bar{\lambda}+2\bar{\mufr})^{R}$, $\bar{\alphaB}^{R}$, and $\bar{M}^{R}$.

The frequency-dependent quality factors $\hat{Q}(\omega) = (Q_{\lambda+2\mufr}, Q_{\mufr},Q_{\alphaB M},Q_{M})$ 
\begin{equation}
\label{eq:qualfacs}
    \hat{Q}_j(\omega) = \frac{\Re (\hat{M}_j)}{\Im (\hat{M}_j)},\quad j=1,\ldots,4,
\end{equation}
where $\hat{M}(\omega) = (\hat{M}_{\lambda+2\mufr}, \hat{M}_{\mufr}, \hat{M}_{\alphaB M}, \hat{M}_{M})$. Finally, we apply a linear optimization procedure \cite{Emmerich} to obtain the anelastic coefficients 
 \begin{eqnarray}
  {\bf Y}^{(n)} = \left(Y_{\lambda+2\mufr}^{(n)}, Y_{\mufr}^{(n)}, Y_{\alphaB M}^{(n)}, Y_{M}^{(n)}\right).
\end{eqnarray}

\subsubsection{Un-relaxed material parameters}\label{sec:unrelaxed}

To obtain the corresponding un-relaxed values, we apply \cite{moczo2014finite}
\begin{eqnarray}
\label{eq:phiu}
\Phi^U_j &=& \Phi^R_j\frac{\sqrt{D_1^2 + D_2^2}+D_1}{2(D_1^2 + D_2^2)},\\
  D_1 &=& 1-\sum_{n=1}^N{\bf Y}_j^{(n)}\frac{\omega_n^2}{\omega_n^2+\omega_r^2},\\
\label{eq:d2}
  D_2 &=& \sum_{n=1}^N{\bf Y}_j^{(n)}\frac{\omega_n\omega_r}{\omega_n^2+\omega_r^2},
\end{eqnarray}
where $j=1,\ldots,4$ and
\begin{eqnarray}
  \Phi^{U/R} = \left(\left(\lambda+2\mufr\right)^{U/R}, {\mufr}^{U/R}, \left({\alphaB M}\right)^{U/R}, {M}^{U/R}\right).
\end{eqnarray}

Finally, the un-relaxed $\lambda$ and the anelastic coefficients $Y_{\lambda}$ are obtained from
\begin{eqnarray}
\label{eq:lambdaR}
  \lambda^{U} &=& \left(\lambda+2\mufr\right)^{U}- 2{\mufr}^{U}, \\
  \label{eq:YlambdaR}
  Y_{\lambda}^{(n)} &=& \frac{\left(\lambda+2\mufr\right)^U Y_{\lambda+2\mufr}^{(n)} - 2\mufr^U Y_{\mufr}^{(n)}}{\left(\lambda + 2\mufr\right)^U-2\mufr^U}.
\end{eqnarray}

\subsection{Viscoelastic wave equation}\label{sec:elastic}

In this study, wave propagation in a purely elastic medium is also considered. Expressed as a second-order system, the elastic wave equation in three dimensions takes the form \cite{aki2002}
\begin{equation}
  \rhoe\pdd{\uue}{t} = \nabla\cdot \boldS,
\end{equation}
where $\uue$ is the displacement, $\rhoe$ is the density, and $\boldS$ is the stress tensor. For the case examined in this study, $\boldS$ may be written in the usual form
\begin{equation}
  \boldS = 2\mue {\bf E} + \lambdae \trace({\bf E}){\bf I}_3,
\end{equation}
where ${\bf E}$ is the solid strain tensor, and $\mue$ and $\lambdae$ are the Lam{\'e} parameters.

To include the attenuation of elastic waves, we approximate the material as a viscoelastic medium. We use the GMB rheological type \cite{moczo2014finite}, and we re-write the solid tensor $\boldS$ components as \cite{7956216}
\begin{align}
\label{eq:Se11_mod}
  S_{11} &= \left(\lambdae + 2\mue\right)^U \epsilon_{11} + \lambdae^U \epsilon_{22} + \lambdae^U \epsilon_{33} - \nonumber \\
  &  
 \sum_{n=1}^N\left(\left(\lambdae + 2\mue\right)^U Y_{\lambdae+2\mue}^{(n)}\theta_{\epsilon_{11}}^{(n)} + \lambdae^U Y_{\lambdae}^{(n)}\theta_{\epsilon_{22}}^{(n)} + \lambdae^U Y_{\lambdae}^{(n)}\theta_{\epsilon_{33}}^{(n)}\right),\\
  S_{22} &=  \lambdae^U \epsilon_{11} + \left(\lambdae + 2\mufr\right)^U\epsilon_{22} + \lambda^U \epsilon_{33} - \nonumber \\
  &  
  \sum_{n=1}^N\left(\lambdae^U Y_{\lambdae}^{(n)}\theta_{\epsilon_{11}}^{(n)} + \left(\lambdae + 2\mue\right)^U Y_{\lambdae+2\mue}^{(n)}\theta_{\epsilon_{22}}^{(n)} + \lambdae^U Y_{\lambdae}^{(n)}\theta_{\epsilon_{33}}^{(n)}\right),
\\
  S_{33} &=  \lambdae^U \epsilon_{11} + \lambdae^U\epsilon_{22} + \left(\lambda + 2\mue\right)^U \epsilon_{33} -  \nonumber \\
  &  
  \sum_{n=1}^N\left(\lambdae^U Y_{\lambdae}^{(n)}\theta_{\epsilon_{11}}^{(n)} + \lambdae^U Y_{\lambdae}^{(n)}\theta_{\epsilon_{22}}^{(n)} + \left(\lambdae + 2\mue\right)^U Y_{\lambdae+2\mue}^{(n)}\theta_{\epsilon_{33}}^{(n)}\right),\\
  S_{12} &= 2\mue^U \epsilon_{12} - 
  \sum_{n=1}^N2\mue^U Y_{\mue}^{(n)}\theta_{\epsilon_{12}}^{(n)},\\
  S_{13} &= 2\mue^U \epsilon_{13} - 
  \sum_{n=1}^N2\mue^U Y_{\mue}^{(n)}\theta_{\epsilon_{13}}^{(n)},\\
\label{eq:Se12_mod}
  S_{23} &= 2\mue^U \epsilon_{23} - 
  \sum_{n=1}^N2\mue^U Y_{\mue}^{(n)}\theta_{\epsilon_{23}}^{(n)}.  
\end{align}
\sloppy Similarly, as in the poroviscoelastic case, also in (\ref{eq:Se11_mod})–(\ref{eq:Se12_mod}), the superscript $U$ denotes the un-relaxed material parameter. The $6N$ memory variables $\bth^{(n)}_{\mathrm{e}} = (\theta_{\epsilon_{11}}^{(n)}, \theta_{\epsilon_{22}}^{(n)}, \theta_{\epsilon_{33}}^{(n)}, \theta_{\epsilon_{12}}^{(n)}, \theta_{\epsilon_{13}}^{(n)}, \theta_{e_{23}}^{(n)})$ also satisfy the ordinary differential equations 
\begin{equation}
\label{eq:thetaeqs_ela}
\frac{\partial \bth_{{\mathrm{e}}, j}^{(n)}}{\partial t} = \omega^{(n)}\left(\bte_{e,j}-\bth_{{\mathrm{e}}, j}^{(n)}\right),\quad j = 1,\ldots,6,
\end{equation}
where $\bte_e = (\epsilon_{11}, \epsilon_{22}, \epsilon_{33}, \epsilon_{12}, \epsilon_{13}, \epsilon_{23})$.

Let us assume that the quality factors ($Q_P,Q_S$) for the pressure and shear waves are known. The optimization procedure \cite{Emmerich} can then be applied to obtain anelastic coefficients $Y_{P}^{(n)}$ and $Y_{S}^{(n)}$ for viscoelastic pressure and shear wave propagation \cite{moczo2014finite}. On the other hand, the anelastic coefficients in terms of  Lam{\'e} parameters $\lambdae$ and $\mue$ are obtained using the transformation  \cite{moczo2014finite}
\begin{eqnarray}
\lambdae^U Y_{\lambdae}^{(n)} + 2\mue^U Y_{\mue}^{(n)} & = & \left(\lambdae+2\mue\right)^U Y_{P}^{(n)},\\
Y_{\mue}^{(n)} & = & Y_{S}^{(n)}.
\end{eqnarray}
Eqs. (\ref{eq:phiu})–(\ref{eq:d2}), given in Section \ref{sec:unrelaxed}, can be used to calculate un-relaxed material parameters with (in this case, $j = 1,2$)
\begin{eqnarray}
  \Phi^{U/R} &=& \left(\left(\lambdae+2\mue\right)^{U/R}, {\mue}^{U/R}\right).
\end{eqnarray}

\section{Discontinuous Galerkin method}\label{sec:dg}

The applied poroviscoelastic–viscoelastic model is expressed as a first-order hyperbolic system \cite{ward2020discontinuous}. We operate in Biot's low-frequency regime (see the physical parameter choices in Section \ref{sec:params} and the discussion in Appendix \ref{sec:biot}), which leads to a stiff partial differential equation system to solve. We apply Godunov splitting \cite{leveque02} to handle the stiffness of the system. Godunov splitting separates the dissipative component from the conservation term at each time step. As done, for example, in \cite{shukla2019}, the stiff part of the system is solved analytically, while the non-stiff part is solved numerically. More precisely, the spatial derivatives of the first-order hyperbolic system are approximated using the nodal DG method \cite{hesthaven_warburton_book}, while time integration is carried out using the third-order Adams–Bashforth scheme \cite{durran91}. 

In the discretized system, we use tetrahedral elements to discretize the model geometry into $N_{el}$ elements. To obtain a strong form, in the DG formulation for each element, we separately multiply the hyperbolic system by a local test function and integrate by parts twice to obtain an element-wise variational formulation. As the numerical flux (i.e., the physical connection between adjacent elements), the upwind scheme \cite{ward2020discontinuous} is used. For a more detailed discussion of the DG method, please refer to \cite{hesthaven_warburton_book} and the references therein. 

In the simulations, the time step $\Delta t$ for the applied Adams–Bashforth scheme is computed from
\begin{equation}
  \label{eq:CFL}
  \Delta t = \min\left(\frac{0.4h^{(\ell)}}{c_{\max}^{(\ell)}(N_b+1)^2}\right),\quad \ell=1,\ldots,N_{el}.
\end{equation}
Here $N_b$ is the basis order. In addition, $c_{\max}^{(\ell)}$ is the maximum wave speed and $h^{(\ell)}$ is the diameter of the largest inscribed sphere for the element $\ell$. 

\subsection*{Implementation}

The current software is written in C/C++ programming language. Communication between different computational CPU nodes is handled using the message passing interface. Currently, the software uses constant order basis functions, so the main reason for the different computational loads on each element comes from the selected physical model (viscoelastic or poroviscoelastic). The computational load between different elements is balanced using the parMetis software in such a way that each element has an individual weight based on its physical model. Finally, the Open Concurrent Compute Abstraction (OCCA) \cite{medina2014occa} library is used for communicating between the CPU and the GPU, as well as for computing on the graphics units. The Puhti environment of the CSC - IT Center for Science \cite{CSC} in Finland is used for the calculations. All the computations shown in the following sections are performed using Puhti's NVIDIA Volta V100 graphics cards.

%%%%%%%%%%%%%%%%%%%%%%
\section*{Acknowledgment}

This work has been supported by the Academy of Finland (the Finnish Centre of Excellence of Inverse Modeling and Imaging) and the Academy of Finland project 321761. The authors also wish to acknowledge the CSC – IT Center for Science, Finland, for generously sharing their computational resources. Special thanks to the Natural Resources Institute Finland (Luke) for sharing information about the sand pool.
%%%%%%%%%%%%%%%%%%%%%%

%\bibliographystyle{abbrv}
%\bibliography{main}

\end{document}